\documentclass[preprintnumbers,showpacs,amsmath,amssymb,twocolumn]{revtex4}

\usepackage{graphicx}
\usepackage{dcolumn}
\usepackage{bm}

\begin{document}

\title{Quantum theory of frequency pulling in the cavity-QED microlaser}

\author{Hyun-Gue Hong}
\author{Kyungwon An}
\address{Department of Physics and Astronomy, Seoul National University, Seoul, 151-742, Korea}

\date{\today}

\begin{abstract}
{The spectrum of the cavity-QED microlaser/micromaser is expected to show distinctive features of the coherent light-matter interaction, which are obscured in the conventional Schawlow-Townes linewidth theory. However, the spectral studies has been limited to resonant atom-cavity interaction so far.
Here we consider the dispersive interaction in the off-resonance case, from which we uncover a quantum frequency pulling effect in the microlaser/micromaser spectrum.
We present a quantum theory of the spectrum which introduces the notion of a frequency-pulling distribution associated with the photon number.
In contrast to the conventional laser, periodic variation of the mean frequency pulling is observed with increasing pump parameter and it is attributed to the strong atom-cavity coupling.
The pulling distribution gives rise to a spectral broadening, which can be dominant over the non-dispersive broadening addressed in the previous works.
We also developed a corresponding semiclassical theory and discuss how the introduction of the frequency shift fits in with the extended quantum theory.}

\end{abstract}

\pacs{42.50Ar, 42.50Nn, 42.50Pq, 42.55Ah}
\maketitle

\section{Introduction}

The microlaser/micromaser \cite{Meschede1985, An1994} operates with a small number of atoms under the strong atom-cavity coupling condition of the cavity quantum electrodynamics (QED). The manifestation of atom-field coherent interaction, which is masked by stochastic averages in the conventional laser, have been observed in the micromaser/microlaser \cite{Benson1994, FangYen2006, Seo2010}. Moreover, its quantum particle property has revealed the nonclassical aspect of the electromagnetic field \cite{Rempe1990, Weidinger1999, Choi2006}.
%
The wave property, on the other hand, which is expected to exhibit the complementary nature still awaits its experimental verification.
In particular, the power spectral density or the first-order coherence is expected to show features \cite{Scully1991} unexplainable by the conventional Schawlow-Townes linewidth theory \cite{Schawlow1958}.
Since the seminal work on the theory of the micromaser spectrum \cite{Scully1991}, more elaborate and refined calculations \cite{Lu1993, Quang1993, Vogel1993, Schieve1997, McGowan1997} have followed.
Indirect measurement of the micromaser spectrum has been proposed by using an atom probe \cite{Brecha1992, Englert1996, Casagrande2003}. In the microlaser operating in the optical region, the spectrum measurement is expected to be rather straightforward.

It should be noted that all those previous works on the micromaser/microlaser spectrum have been restricted to the atom-cavity resonance condition, and thus they cannot be applied to off-resonance cases by simple modifications
because dispersive interactions can induce a frequency shift, resulting in a qualitatively different spectral lineshape.
Even in the conventional laser, for example,  the phenomena called {\it frequency pulling} or {\it mode pulling} makes the true laser oscillation frequency shifted from a passive cavity resonance frequency in the presence of a frequency detuning between the gain medium and the cavity.
It is thus natural to ask how the spectrum changes as we remove the constraint of resonance over the whole tuning range of the microlaser.

Here we present an extended quantum theory of the microlaser spectrum including off-resonance cases.
Our theory predicts a quantum frequency pulling effect, a novel feature not found in the conventional laser.
Moreover, the amount of the frequency pulling strongly depends on the pumping parameter owing to the strong atom-cavity coupling.
In the standard laser theory of Lamb and Scully \cite{ScullyBook}, in contrast, the amount of the pulling is solely determined by the linewidths of the atomic and the cavity resonances independent of a pumping parameter.
In particular, the unitary evolution of the atomic state gives rise to a periodic characteristic in the pulling.
The analysis given here is also applicable to the microwave counterpart as long as thermal microwave photons are negligible.

The quantum frequency pulling to be discussed below is indexed by the photon number, the uncertainty of which leads to spreading in the frequency shift or a spectral broadening. This mechanism of dispersive broadening has never been addressed so far although it can easily exceed the nondispersive contribution in the spectral width under the present experimental conditions of the microlaser.
Only microscopic lasers like the microlaser/micromaser are adequate to study this quantum dispersive effect:
in the conventional laser, the differential pulling by one-photon increment/decrement is negligible and thus not a noticeable effect.
We note that laser oscillation frequency shift with respect to the cavity has been discussed in the socalled mazer \cite{Schroder1997}, a maser pumped by the matter wave, and in the trapping state spectrum \cite{Lu1993}, but not in a general context as in the present paper.
To help gain an insight on the frequency pulling effect, we have also present a corresponding semiclassical theory.

This paper is composed of the quantum theory in Sec.\ \ref{sec:quantum} and the semiclassical theory in Sec.\ \ref{sec:semiclassic}. The quantum theory is based on the quantum regression (Sec.\ \ref{subsec:qreg}) and the master equation (Sec.\ \ref{subsec:master}). Central analytic formulae are derived in Sec.\ \ref{subsec:eigenvalue} and Appendix \ref{subsec:formulae}.
After examining the physical meaning of the result of the quantum theory and its conventional laser limit (Sec.\ \ref{subsec:meaning}), typical spectral lineshapes (Sec.\ref{subsec:lineshape}) and the dependence on detuning (Sec.\ref{subsec:detuning}) and pumping (Sec.\ref{subsec:pumping}) are investigated. The semiclassical theory presented in Sec.\ \ref{subsec:msequation} is based on the Maxwell-Schr\"{o}dinger equation and is compared with the quantum theory in Sec.\ \ref{subsec:ndelta}. The correction made by a newly-introduced frequency variable in the semiclassical theory is elucidated
by using a graphical method in Appendix \ref{subsec:graphic}.

\section{Quantum theory}\label{sec:quantum}

In this section we generalize the master-equation-based theory of the micromaser spectrum \cite{Scully1991} in order to include the effect of atom-cavity detuning. The key expressions relating steady-state variables are derived analytically. The numerical solution of the master equation is used to determine these steady-state variables and, in turn, to perform exact calculation of the quantities related to the frequency pulling. The quantum distribution of the frequency pulling is naturally introduced along this line of analysis.

\subsection{First-order correlation function}\label{subsec:qreg}

The spectrum of an optical field is obtained by taking the Fourier transform of its first-order correlation function which is written as
\setlength\arraycolsep{2pt}
\begin{eqnarray}
\langle a^{\dag}(t)a(0)\rangle={\textrm{tr}}_{F\oplus R}\left[\chi(0)a^{\dag}(t)a(0)\right],
\end{eqnarray}
where $\chi$ is the density
operator of the field(F)-reservoir(R) system and $a^\dagger$ and $a$ are the creation and annihilation operators, respectively, of the laser field. The time origin $t=0$ here is set to an arbitrary moment at which the steady state of the system is achieved. We
have knowledge on the evolution of the field subsystem,
which is described by
\begin{eqnarray}
\rho(t)={\textrm{tr}}_{R}\left[\chi(t)\right]
\end{eqnarray}
via the master equation
\begin{eqnarray} {\dot \rho}(t)={\cal L}\rho(t),
\end{eqnarray}
the detailed form of which will be described below. The time evolution of the total system is given by
$U(t)$ such that $a^{\dag}(t)=U^{\dag}(t)a^{\dag}(0)U(t)$.
We proceed as
\begin{eqnarray}
\langle a^{\dag}(t)a(0)\rangle&=&{\textrm{tr}}_{F\oplus
R}\left[\chi(0)U^{\dag}(t)a^{\dag}(0)U(t)a(0)\right]\nonumber\\&=&{\textrm{tr}}_{F}[a^{\dag}(0){\textrm{tr}}_{R}
\left[U(t)a(0)\chi(0)U^{\dag}(t)\right]]\nonumber\\&\equiv&{\textrm{tr}}_{F}[a^{\dag}(0)\tilde{\rho}(t)],
\end{eqnarray} to give \begin{equation}
\langle a^{\dag}(t)a(0)\rangle=\langle
a^{\dag}\rangle_{\tilde{\rho}(t)}.
\end{equation}
Therefore, $\langle a^{\dag}(t)a(0)\rangle$ is the expectation value of the field amplitude subject to the transformed density matrix
$\tilde{\rho}(t)={\textrm{tr}}_{R} \left[U(t)a(0)\chi(0)U^{\dag}(t)\right]$.
According to the quantum regression theorem \cite{ScullyBook}, the equation of motion for $\tilde{\rho}(t)$ is given by that of the $\rho(t)$,
that is,
\begin{equation} \dot
{\tilde{\rho}}(t)={\cal L}\tilde{\rho}(t).\label{eq:rhoDE}
\end{equation} In the photon number basis the first-order coherence function is related to the off-diagonal element of the density matrix ${\tilde{\rho}}_{n,n+1}\equiv{\tilde {\rho}^{(1)}}_n$ as
\begin{equation}
\langle a^{\dag}(t)a(0)\rangle=\sum_n\sqrt{n+1}\tilde{\rho}^{(1)}_n(t),
\label{eq:aa}
\end{equation}
where $\tilde{\rho}^{(1)}_n(t)$ is the solution of Eq.\ (\ref{eq:rhoDE}).
The initial value of the transformed density matrix is given by
\begin{equation}
\tilde{\rho}(0)={\textrm{tr}}_{R}\left[a(0)\chi(0)\right]=a(0)\rho(0),
\end{equation}
which is in the photon number basis
\begin{equation} \langle
n|\tilde{\rho}(0)|n+1
\rangle=\sqrt{n+1}\langle n+1|\rho(0)| n+1
\rangle \end{equation}
with all other terms being zero.

\subsection{Master equation for  $\rho^{(1)}_n$}\label{subsec:master}

The next task is to find the time evolution of $\rho_n^{(1)}=\rho_{n,n+1}$. The master equation is written by considering the change of the field due to the coherent interaction with the initially inverted atoms and the incoherent decay. We assume the tophat interaction model in which the atom-cavity interaction parameterized by the constant atom-cavity coupling $g$ persists for a fixed amount of time $\tau$ for all atoms that traverse the cavity mode. The mean number of atoms inside the interaction region is designated by $N$. The realistic modification such as the spatial variation of the cavity mode and the inhomogeneous distribution of the interaction time can be readily included in the numerical calculation.

The coarse-grained master equation for the microlaser, written in the photon number basis, reads \cite{ScullyBook}
\begin{eqnarray}
{\dot \rho}_{nm}=a_{nm}\rho_{n,m}+b_{nm}\rho_{n+1,m+1}+c_{nm}\rho_{n-1,m-1}
\label{eq:master1}\end{eqnarray}
where
\begin{eqnarray}
a_{nm}&=&r_a\biggl[\cos\left(\frac{\Omega_n\tau}{2}\right)\cos\left(\frac{\Omega_m\tau}{2}\right) \nonumber \\
& &+\frac{\Delta'^2}{\Omega_n\Omega_m}\sin\left(\frac{\Omega_n\tau}{2}\right)\sin\left(\frac{\Omega_m\tau}{2}\right)-1\nonumber\\
& &-i\frac{\Delta'}{\Omega_m}\cos\left(\frac{\Omega_n\tau}{2}\right)\sin\left(\frac{\Omega_m\tau}{2}\right) \nonumber \\
& & +i\frac{\Delta'}{\Omega_n}\cos\left(\frac{\Omega_m\tau}{2}\right)\sin\left(\frac{\Omega_n\tau}{2}\right)\biggr]-\gamma_c(n+m), \nonumber\\
& &
\label{eq:master-a}\end{eqnarray}
\begin{eqnarray}
b_{nm}=2\gamma_c\sqrt{(n+1)(m+1)},
\label{eq:master-b}\end{eqnarray}
\begin{eqnarray}
c_{nm}=r_a\biggr[\frac{4g^2\sqrt{nm}}{\Omega_{n-1}\Omega_{m-1}}\sin\left(\frac{\Omega_{n-1}\tau}{2}\right)\sin\left(\frac{\Omega_{m-1}\tau}{2}\right) \biggr],\nonumber\\
\label{eq:master-c}\end{eqnarray}
with the {\it field}-atom detuning $\Delta'=\omega-\omega_0$, the $n$-photon Rabi frequency $\Omega_n=\sqrt{4g^2(n+1)+\Delta'^2}$, the injection rate of the atomic beam $r_a=N/\tau$, and the cavity decay rate $\gamma_c$ (a half width). Note that the frequency of the laser field $\omega$ does not necessarily coincide with the passive cavity frequency $\omega_c$ if we are open to possible frequency shift. In that sense we reserve $\Delta$ for the passive cavity detuning $\Delta=\omega_c-\omega_0$. Since we are interested in the steady-state operation, in which the frequency of the field is already shifted, the master equation is written in terms of the resulting shifted detuning $\Delta'$. 
The introduction of $\Delta'$ and the corresponding modification of $\Omega_n$ in Eqs.\ (\ref{eq:master-a})-(\ref{eq:master-c}) are the key changes made to the quantum theory of Ref.\ \cite{Scully1991}.
The imaginary part of Eq.\ (\ref{eq:master1}), which appears only in the presence of a nonzero detuning, already suggests the introduction of  frequency pulling since the steady state ${\dot \rho}_{nm}=0$ cannot be achieved without the frequency shifting transformation such as $\rho'_{nm}(t)=\rho_{nm}(t) e^{i(n-m)(\Delta'-\Delta)t}$.

\subsection{Complex eigenvalue of $\rho_n^{(1)}$ and the emergence of  frequency pulling}\label{subsec:eigenvalue}

The equation of motion for the off-diagonal element $\rho_n^{(1)}$ appears as
\begin{widetext}\begin{eqnarray}
{\dot \rho^{(1)}}_n&=&-\frac{1}{2}\mu_n\rho_n^{(1)}+r_a\frac{4g^2\sqrt{n(n+1)}}{\Omega_{n-1}\Omega_n}\sin\left(\frac{\Omega_{n-1}\tau}{2}\right)\sin\left(\frac{\Omega_{n}\tau}{2}\right)\rho_{n-1}^{(1)}-2\gamma_c\sqrt{n(n+1)}\rho^{(1)}_n
\nonumber\\&&-r_a\frac{4g^2\sqrt{(n+1)(n+2)}}{\Omega_{n}\Omega_{n+1}}\sin\left(\frac{\Omega_n\tau}{2}\right)\sin\left(\frac{\Omega_{n+1}\tau}{2}\right)\rho^{(1)}_n+2\gamma_c\sqrt{(n+1)(n+2)}\rho_{n+1}^{(1)}
\label{coarseMaster1} \end{eqnarray} where we defined
\begin{eqnarray}
-\frac{1}{2}\mu_n&\equiv&
r_a\biggl\lbrace\cos\left(\frac{\Omega_n\tau}{2}\right)\cos\left(\frac{\Omega_{n+1}\tau}{2}\right)
+\frac{\Delta'^2}{\Omega_n\Omega_{n+1}}\sin\left(\frac{\Omega_n\tau}{2}\right)\sin\left(\frac{\Omega_{n+1}\tau}{2}\right)
-i\frac{\Delta'}{\Omega_{n+1}}\cos\left(\frac{\Omega_n\tau}{2}\right)\sin\left(\frac{\Omega_{n+1}\tau}{2}\right)
\nonumber\\&&+i\frac{\Delta'}{\Omega_n}\cos\left(\frac{\Omega_{n+1}\tau}{2}\right)\sin\left(\frac{\Omega_n\tau}{2}\right)
+\frac{4g^2\sqrt{(n+1)(n+2)}}{\Omega_{n}\Omega_{n+1}}\sin\left(\frac{\Omega_n\tau}{2}\right)\sin\left(\frac{\Omega_{n+1}\tau}{2}\right)-1\biggr\rbrace+2\gamma_c\sqrt{n(n+1)}
\nonumber\\&&-\gamma_c(2n+1),
\label{eq:mu_n}\end{eqnarray}\end{widetext}
and we rearranged the terms in order to employ the detailed balance between
$2\gamma_c\sqrt{(n+1)(n+2)}\rho_{n+1}^{(1)}$ and
$r_a\frac{4g^2\sqrt{(n+1)(n+2)}}{\Omega_{n}\Omega_{n+1}}\sin\left(\frac{\Omega_n\tau}{2}\right)\sin\left(\frac{\Omega_{n+1}\tau}{2}\right)\rho^{(1)}_n$ in the steady state \cite{Scully1991}.
Such a balance guides $\rho^{(1)}_n$ to decay exponentially with a single time constant $\frac{1}{2}\mu_n$.
The detailed balance holds for most of the parameter space except the trapping state condition \cite{Weidinger1999, Meystre1988}, at which the truncation of the Hilbert state occurs.
We note that the single-decay ansatz is known to give more or less an inaccurate value of the linewidth for the trapping state \cite{McGowan1997}, and the correct spectrum often comes up with a nontrivial lineshape \cite{Lu1993}. We exclude this singular case in our discussion for simplicity.
As noted in the definition of $\Delta'$ and $\Delta$ in Sec.\ \ref{subsec:master}, $\Delta'=\Delta+\langle \delta_n\rangle$ is understood in the steady state.

The corresponding field correlation function is then obtained from Eq.\ (\ref{eq:aa}) with $t>0$ as
\begin{eqnarray}
\langle
a^{\dag}(t)a(0)\rangle&=&\sum_n\sqrt{n+1}\tilde{\rho}^{(1)}_n(0)e^{-\frac{1}{2}\mu_n
t}\nonumber\\&=&\sum_n(n+1)p_{n+1}e^{-\frac{1}{2}\mu_n
t}\nonumber\\&\equiv&\sum_n(n+1)p_{n+1}e^{-D_nt+i\delta_nt}\label{eq:correlation1}\end{eqnarray}
where $D_n$ and $-\delta_n$ are defined as the real and imaginary parts of $\frac{1}{2}\mu_n$, respectively, and $p_{n}$ is the steady-state photon number distribution which can be obtained by the standard theory of the micromaser \cite{Filipowicz1986}.
The optical spectrum obtained by
\begin{equation}
S(\nu-\omega_c)=\frac{1}{\pi}{\rm Re}\int ^{\infty}_{0}\langle a^{\dag}(t)a(0)\rangle e^{-i(\nu-\omega_c)t}dt\label{eq:wienerFT}
\end{equation}
is therefore the sum of individual Lorentzians centered at $\nu=\omega_c+\delta_n$ with a half linewidth $D_n$. Here $D_n$ is identified with a phase diffusion constant at the fixed oscillation frequency while $\delta_n$ is interpreted as a frequency shift. We note both $D_n$ and $\delta_n$ depend on $n$, and thereby have distributions associated with the photon number. While the phase diffusion $D_n$ represents a spectral broadening by itself, the frequency shift $\delta_n$ gives rise to an additional broadening by spreading out the center frequencies of the individual Lorentzians. Note that the contribution from each Lorentzian is weighted by $np_n$.

In the limit of $n\gg1$, the real part $D_n$ is further simplified as
\begin{eqnarray}
D_n&\simeq&2r_a\sin^2\left(\frac{g^2\tau}{2\Omega_n} \right)+\frac{\gamma_c}{4n}
\end{eqnarray}
as shown in Appendix \ref{subsec:formulae}. This result appears as a simple generalization of the on-resonance result \cite{Scully1991} by replacing $2g\sqrt{n+1}$ in the denominator of the sine term with $\Omega_n$.

Let us examine the imaginary part $\delta_n$ which has never been addressed so far in the literature. In the same limit of $n\gg1$, it is simplified as (see Appendix \ref{subsec:formulae})
\begin{eqnarray}
\delta_n&\simeq&-r_a\frac{g^2\Delta'\tau}{\Omega_n^2}\biggl[1-\frac{\sin(\Omega_n\tau)}{\Omega_n\tau} \biggr].
\label{eq:pulling1}
\end{eqnarray}
Notice that, on resonance ($\Delta=0$), the frequency pulling $\delta_n$ should vanish to satisfy Eq.\ (\ref{eq:pulling1}). The frequency pulling is truly a result of dispersive atom-cavity interaction, which induces an additional optical path length other than that of the cold cavity. The overall minus sign indicates that the sign of $\delta_n$ is opposite to that of the detuning so that the oscillation
frequency of the microlaser is pulled towards the atomic resonance.

Since $\Omega_n$ in the right-hand-side of Eq.\ (\ref{eq:pulling1}) also contains $\langle \delta_n\rangle$ in the effective detuning, Eq.\ (\ref{eq:pulling1}) can not be written in a closed form equation for $\delta_n$. Although some of its properties can be inferred from an analytic form as we will do below, the exact calculation of the frequency pulling is presented by numerical means mostly with the parameters relevant to the microlaser experiment of Refs.\ \cite{FangYen2006, Choi2006} throughout this paper. In the numerical calculation the equivalent Gaussian mode function is employed to be more realistic.

\subsection{Physical meaning and the conventional laser limit of the frequency pulling}\label{subsec:meaning}

\begin{figure}[t] \centering
\includegraphics[angle=0,width=3.4in]{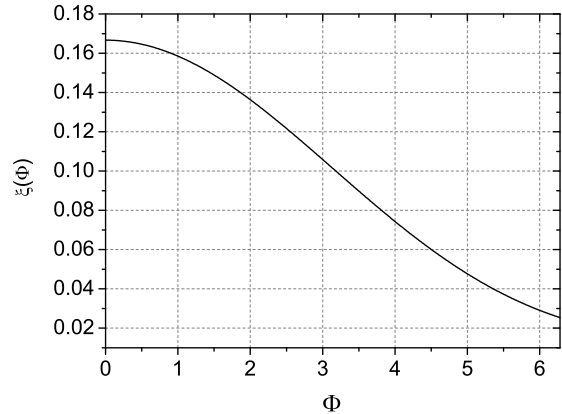}
\caption{The characteristic function $\xi(\Phi)$ representing the Rabi angle dependence of the frequency pulling.} \label{fig:xiFunction}
\end{figure}

Before we proceed with systematic numerical calculations, let us appreciate the physical meaning of Eq.\ (\ref{eq:pulling1}). The mean value of the frequency pulling is approximately equal to $\delta_n$ with $n\simeq\langle n \rangle$. We can rearrange the terms  in Eq.\ (\ref{eq:pulling1}) as
\begin{equation}
\langle\delta_n\rangle=-\frac{\gamma_c\Delta'}{\tau^{-1}}\times
\frac{N}{N_{th}}\times\xi(\Phi)\label{eq:pulling2}\end{equation}
where the characteristic function $\xi(\Phi)$ (plotted in Fig.\ \ref{fig:xiFunction}) is defined as
\begin{equation}
\xi(\Phi)=\left(1-\frac{\sin\Phi}{\Phi}
\right)/\Phi^2.
\end{equation}
Here we used the threshold atom number $N_{th}\equiv 2\gamma_c/(g^2\tau)$ and the semiclassical Rabi angle $\Phi\equiv\sqrt{4g^2\langle n\rangle+\Delta'^2}\tau\equiv\Omega\tau$ assuming $\langle n\rangle\gg1$.

The first multiplicative factor in the right-hand-side of Eq.\ (\ref{eq:pulling2}) corresponds to $\gamma_c \Delta/2\gamma_p$, the frequency pulling in the conventional laser \cite{ScullyBook}, if we identify the dephasing rate of the medium $\gamma_p$ with the transit-time bandwidth $(2\tau)^{-1}$. It has a constant value determined by the relevant bandwidths of the cavity and the gain medium, independent of the pumping rate. In the case of the microlaser we have additional multiplicative factors in Eq.\ (\ref{eq:pulling2}).

The factor $N/N_{th}$, equivalent to the effective pumping parameter often denoted by $\theta^2=Ng^2\tau/2\gamma_c$ \cite{Filipowicz1986},
shows the increase of the frequency shift by pumping
while $\xi(\Phi)$ term indicates the reduction resulting from a Rabi angle associated with the growing laser intensity.
In the conventional laser these opposite tendencies balance each other to result in a constant amount of the pulling independent of the pumping. In other words, the gain saturation approximately leads to $N\propto\langle n\rangle\propto \Phi^2/g^2$ above the threshold while $\xi(\Phi)\simeq\Phi^{-2}$ makes $N\xi(\Phi)$ constant.

In the microlaser, on the other hand, the gain is oscillatory with $\langle n \rangle$ so that we encounter the reduction of gain well above the threshold and the corresponding saturation of $\langle n \rangle$ as a function of $N$ [see Fig.\ \ref{fig:pumping2}(a) below]. The pumping by the atomic flux $N$ is thus translated into the increase of the amount of pulling.

The other pumping channel, the coupling constant $g$, brings the mean photon number down when it is increased, and the corresponding variation of $\Phi^2$ with respect to $g$ is rather small. Thus $g^2$ behavior contained in $N/N_{th}$ term is dominant also in this case. The quadratic dependence comes from the feedback mechanism inherent in the laser amplification process, in which the atomic polarization induced by the cavity field via coupling $g$ drives the cavity field again by the same coupling. 

The notable exception of the general tendency of $Ng^2$-type increase in the frequency pulling happens at the socalled multiple thresholds, which are associated with the periodic Rabi flopping \cite{FangYen2006}. In this case, $\langle n\rangle$ changes discontinuously, and so do $\Phi$ and $\xi(\Phi)$ resulting in the periodic rise and fall of the pulling with $\theta^2$. Thus the microlaser operation is segmented into branch solutions associated with multiple thresholds.
 
The discontinuous change of the system can be explained in terms of the coherent evolution of the atomic state. As the system approaches the saturation regime out of the linear regime of each branch solution, the precession of the Bloch vector representing the internal state of the gain medium reduces the emission probability of photon from each atom while the phase shift per atom grows. At the transition to a higher branch solution, the system is brought to a new steady state by making the precession angle of the Bloch vector exceed $2\pi$. In the middle of such interaction the atomic state traces back to its initial state and starts the evolution all over again. In doing so, the emission probability is boosted up and results in a sudden rise of the mean photon number -- also known as a quantum jump \cite{Benson1994} -- while the phase shift per atom diminishes. Hence, upon the multiple thresholds, the microlaser behaves in a counter-intuitive manner in the sense that it becomes less refractive with a more dense medium of atoms. This nontrivial pumping dependence based on the strong atom-cavity coupling is one of the distinctive features of the microlaser. The argument given here will be examined in detail by numerical calculations in Sec.\ \ref{subsec:pumping}.

The result of the conventional laser can be recovered if we average the microlaser result over a fictitious dwell time distribution given by $P(\tau)=2\gamma_p e^{-2\gamma_p\tau}$ \cite{ScullyBook}, which spoils the well-defined atom-cavity coupling by introducing stochastic interaction time. Using $\int_{0}^{\infty}\tau e^{-2\gamma_p\tau}\left(1-\frac{\sin\Omega\tau}{\Omega\tau}\right)d\tau=\frac{\Omega^2}{4\gamma_p^2(4\gamma_p^2+\Omega^2)}$ and assuming the pulling to be small compared to the detuning ($\Delta'\simeq\Delta$) we have
\begin{equation}
\langle \delta_n \rangle\simeq -r_a\frac{g^2\Delta}{2\gamma_p(4\gamma_p^2+\Omega^2)}=-\frac{\Delta}{2\gamma_p}\gamma_c
\end{equation}
where the last equality employs the steady state condition of the conventional laser $\frac{2r_ag^2}{4\gamma_p^2+\Omega^2}=2\gamma_c$ \cite{ScullyBook} .

\subsection{Spectral lineshape}\label{subsec:lineshape}

\begin{figure} \centering
\includegraphics[width=3.4in]{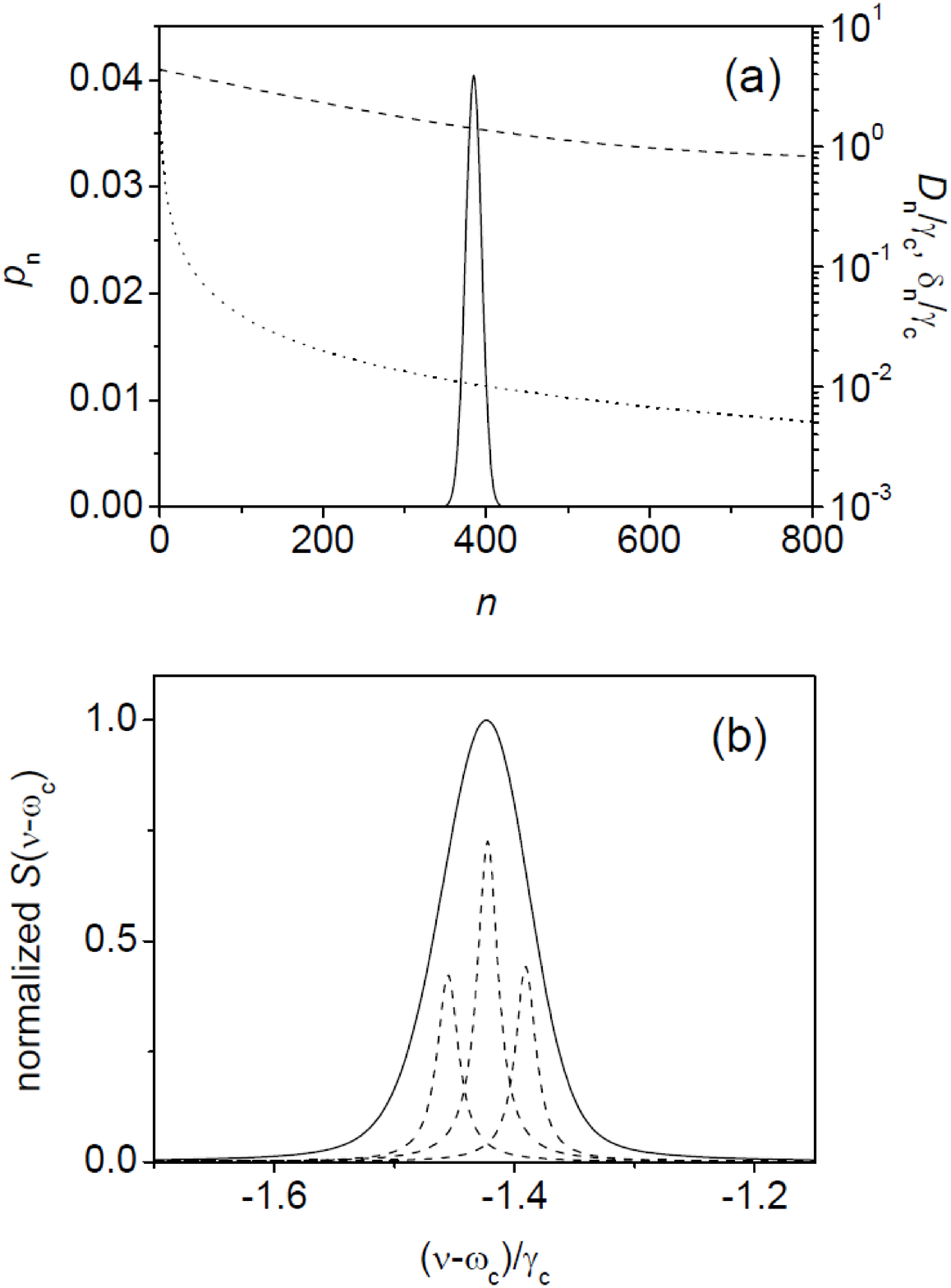}
\caption{(a) The photon number distribution $p_n$ (solid curve), the corresponding $D_n$ (dotted curve) and $\delta_n$ (dashed curve) calculated with $g\tau=0.124$ and $\gamma_c\tau=0.049$, corresponding to the experimental parameters of Refs.\ \cite{Choi2006, FangYen2006}.
The tuning parameters are fixed at $N=100$ and $\Delta/\gamma_c=14.76$. (b) The normalized spectrum as a result of superposing the Lorentzian curves corresponding to different $n$'s. The major spectral energy is confined in a frequency range lower than $\nu=\omega_c$. The constituent Lorentzian curves with $\Delta n=\pm 10$ are depicted in dashed curves for comparison. The spectrum is normalized to a unity height.} \label{fig:lineshape1}
\end{figure}

The overall lineshape is given by a sum of Lorentzian curves spread by the photon number distribution $np_n$. The amount of the spread $\delta_n$ as well as $p_n$ depends on the detuning $\Delta$, thus we expect a significant change of the lineshape as we tune the cavity. The overall spectral shift from the passive cavity resonance is given by $\delta_{\bar{n}}$ with $\bar{n}$ corresponding to a predominant $p_{\bar{n}}$. To obtain the detailed spectral distribution we have to take into account both the intrinsic linewidth $D_n$ and the distribution of the shift $\delta_n$.

Let us for now assume a single-peaked $p_n$ as is the case for most of the microlaser parameters. To a good approximation, $p_n$ with a large $\langle n \rangle$ is well described by the Gaussian curve. The constituent Lorentzian curves have little variation in their widths $D_n$ within the range of substantial probability $p_n$ as in Fig.\ \ref{fig:lineshape1}(a). Thus the spectral lineshape is approximately given by a Voigt integral centered around $\omega_c+\langle \delta_n \rangle$ as in Fig.\ \ref{fig:lineshape1}(b). Of course the lineshape exactly reduces to a Lorentzian on resonance. The pulling-induced broadening exceeds the linewidth of any constituent Lorentzian even with a detuning of several times of $\gamma_c$ away from the atomic resonance, thus becomes the dominant source of decoherence for a large part of the tuning range of the microlaser.

One may be tempted to use this $n$-dependent shift to obtain information on the photon number distribution, which is possible if the differential pulling is much larger than the individual linewidth $D_n$. Unfortunately, we could not identify a parameter set satisfying this condition
because any parameter sets we tried not only increased the differential pulling but also the individual linewidth as well.

\begin{figure} \centering
\includegraphics[width=3.4in]{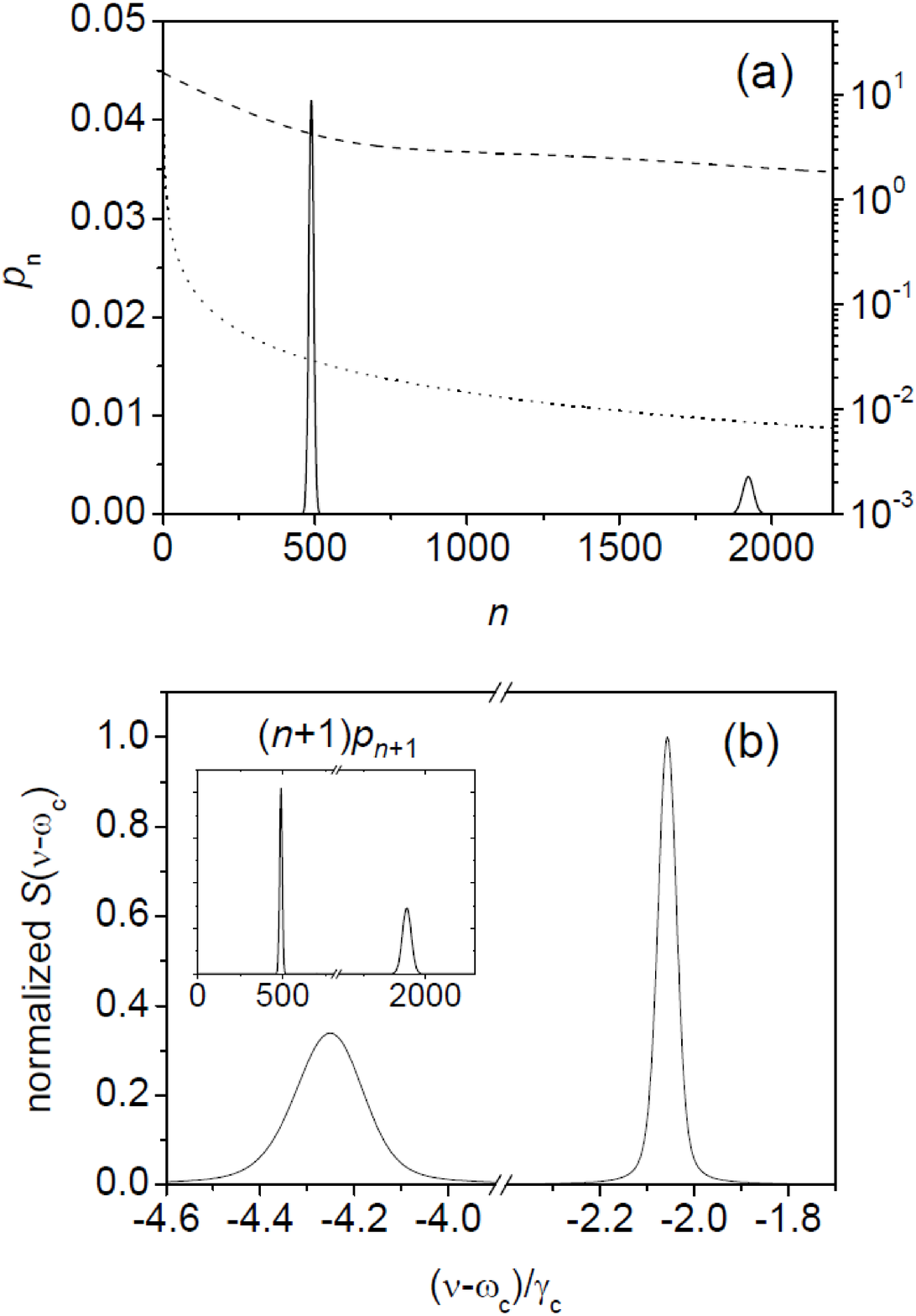}
\caption{Plot of the same quantities as in Fig.\ \ref{fig:lineshape1} except for the tuning parameters set to a multiple-threshold transition point, $N=367$ and $\Delta/\gamma_c=16.61$. The inset in (b) shows the weight function $(n+1)p_{n+1}$.} \label{fig:lineshape2}
\end{figure}

However, the photon number distribution can do a significant role in the spectrum at some particular operating points. It is well known that the photon number distribution is not single peaked if the system undergoes any of the multiple thresholds as shown in Fig.\ \ref{fig:lineshape2}(a) \cite{Benson1994, FangYen2006}. Although $\delta_n$ is a slowly varying function of $n$, the large separation between the two possible solutions of $p_n$ is enough to separate the spectral peaks as depicted in Fig.\ref{fig:lineshape2}(b).
In this case, 
the narrower photon number distribution in the lower branch solution near $n\simeq500$ corresponds to a broader spectrum near $\nu-\omega_c\simeq-4.25$.
This is because the higher branch solution, despite the relatively broad $p_n$, generally corresponds to an order-of-magnitude sharper spectral peak $D_n$ as well as much smaller $\delta_n$. In other words, the spectral energy is redistributed according to $\delta_n$ and $D_n$ while the ratio of the integrated area under each peak of $np_{n}$, about 1.2:1 as shown in the inset of Fig.\ \ref{fig:lineshape2}(b), is preserved in the spectral domain.
In reverse, one should also be able to obtain the binary $p_n$ distribution from the observed dual-peak spectrum as long as $D_n$ and $\delta_n$ in each peak of $p_n$ are approximately constant.

\subsection{Detuning dependence}\label{subsec:detuning}

\begin{figure} \centering
\includegraphics[width=3.4in]{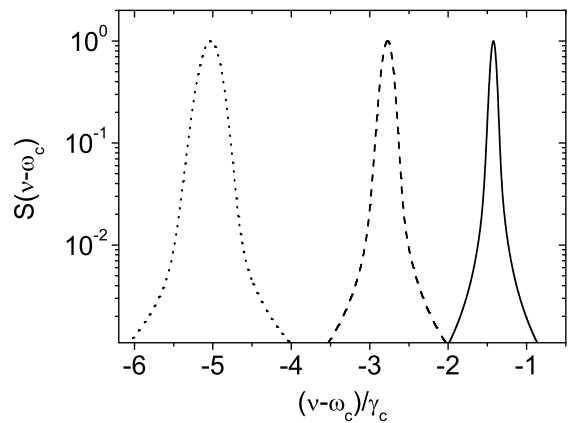}
\caption{The spectral lineshape depending on the detuning $\Delta/\gamma_c=$11.87 (solid), 23.88 (dashed), 48.29 (dotted). The system parameters are $g\tau=0.124$, $\gamma_c\tau=0.049$, and $N=100$.} \label{fig:detuning1}
\end{figure}

The detuning dependence of the spectrum confirms that the oscillation frequency of the microlaser is indeed {\it pulled} toward the atomic line center as in the conventional laser. The magnitude of the effective detuning $\Delta'$ is always smaller than that of the passive detuning $\Delta$. The spectra calculated with three different detunings are depicted in Fig.\ \ref{fig:detuning1} with respect to the passive cavity frequency. We note that the center frequency of the spectra moves the farther away from $\omega_c$ for the larger detuning. The lineshape transforms from a Lorentzian-like curve to a Gaussian-like curve as the frequency excursion induced by $\delta_n$ gets larger than the individual width $D_n$.

\begin{figure} \centering
\includegraphics[width=3.4in]{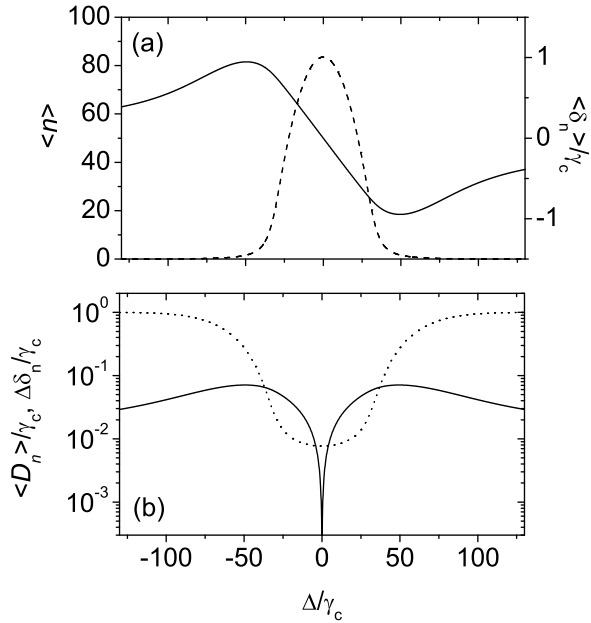}
\caption{(a) The mean photon number $\langle n\rangle$ (dashed) and the mean frequency pulling $\langle \delta_n \rangle/\gamma_c$ (solid). (b) The spectral broadening $\langle D_n\rangle/\gamma_c$ (dotted) and $\Delta\delta_n/\gamma_c$ (solid) as the passive cavity frequency is tuned to near resonance. Plotted with $g\tau=0.124$, $\gamma_c\tau=0.049$, and $N=10$.} \label{fig:detuning2}
\end{figure}

Let us now examine the mean $\langle \delta_n\rangle$ and the variance $\Delta\delta_n$ of the pulling distribution systematically. Before the microlaser reaches the second threshold, {\em i.e.}, within the first branch solution, the mean spectral shift $\langle \delta_n \rangle$ as a function of $\Delta$ follows a familiar dispersion curve [Fig.\ \ref{fig:detuning2}(a)]. The spectral broadening $\Delta\delta_n$ is accordingly maximized at the detuning which maximizes $\langle \delta_n \rangle$  
[Fig.\ \ref{fig:detuning2}(b)]. There we used the standard deviation of $\delta_n$ as a measure of the linewidth, that is, $\Delta\delta_n\equiv\sqrt{\langle \delta^2_n\rangle-\langle \delta_n\rangle^2}$. We note that the relative contribution of $D_n$ becomes significant on resonance, for which the contribution of $\delta_n$ just vanishes. The contribution by $D_n$ is also dominant at far-off resonance, which corresponds to a below-threshold regime; since the field is just the atomic emission, not a lasing field with feedback, filtered by the cavity, we confirm $\langle D_n\rangle\simeq\gamma_c$ there.

\begin{figure} \centering
\includegraphics[width=3.4in]{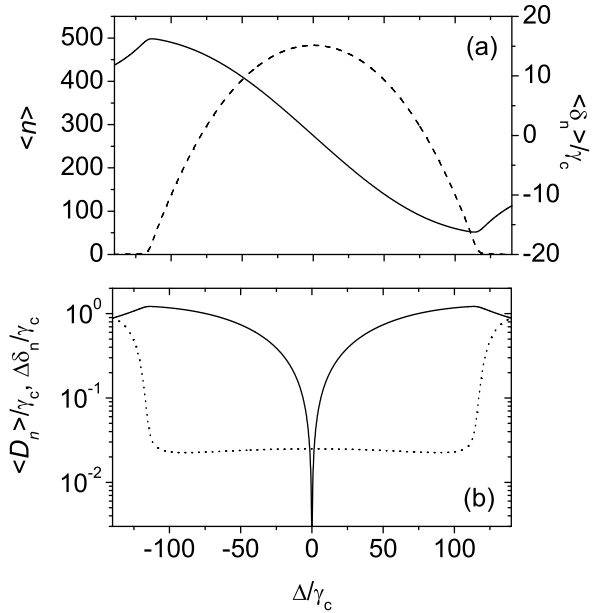}
\caption{Same as Fig.\ \ref{fig:detuning2} except for $N=300$ to reach the saturation regime.} \label{fig:detuning3}
\end{figure}

In the saturation regime within each branch solution, on the other hand, the curve $\langle \delta_n\rangle$ vs. $\Delta$  turns into a power-broadened dispersion curve (Fig.\ \ref{fig:detuning3}).
The contribution of the dispersive broadening becomes dominant almost over the whole tuning range of the microlaser. Thus the spectrum cannot be explained without the notion of the frequency pulling.

\begin{figure} \centering
\includegraphics[width=3.4in]{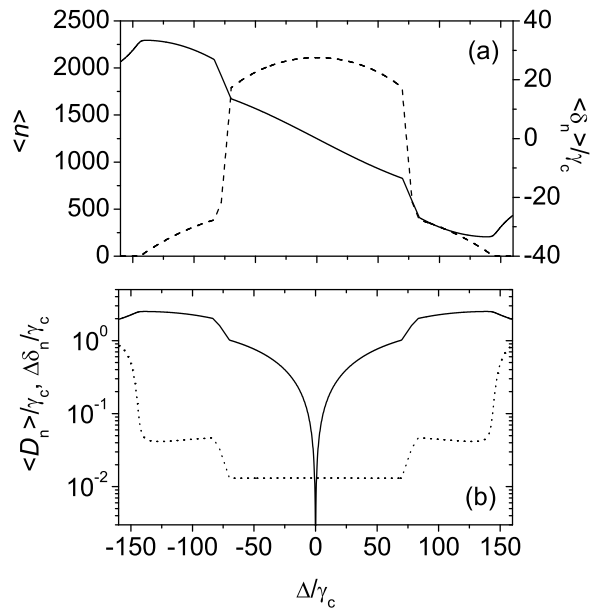}
\caption{Same as Fig.\ \ref{fig:detuning2} except for $N=700$ to reach the second branch solution near resonance.} \label{fig:detuning4}
\end{figure}

An extraordinary dispersion curve can be obtained if we pump the system (see also Sec.\ \ref{subsec:pumping} below) harder so that more than one branch solution can be attained within the cavity-tuning range. The mean photon number or the output intensity of the microlaser as a function of the detuning is known to form a set of discontinuous fragments of individual branch solutions \cite{FangYen2006, Hong2009} as shown in Fig.\ \ref{fig:detuning4}(a). At the very point of the transition where the fragments are connected, the system is bistable as discussed above in Fig.\ \ref{fig:lineshape2}. Note that the higher branch, which occurs at the smaller detuning, has the larger mean photon number. The detuning curve of $\langle \delta_n\rangle$ is also a discontinuous stepwise dispersion curve. The amount of the pulling $\langle \delta_n\rangle$ is, however, smaller for the higher branch in Fig.\ \ref{fig:detuning4}(a). This opposed behavior of $\langle n \rangle$ and $\langle \delta_n \rangle$ upon the multiple thresholds is in agreement with the qualitative explanation previously given in Sec.\ \ref{subsec:meaning}.

\begin{figure} \centering
\includegraphics[width=3.4in]{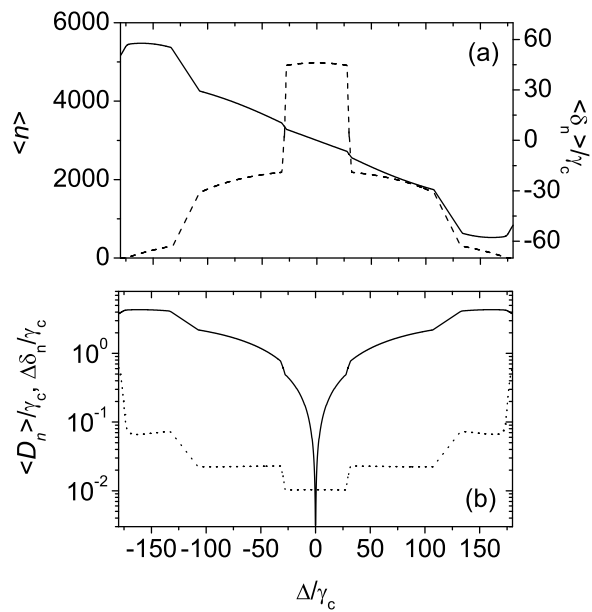}
\caption{Same as Fig.\ \ref{fig:detuning2} except for $N=1300$ to reach the third branch solution near resonance.} \label{fig:detuning5}
\end{figure}

\subsection{Pumping dependence}\label{subsec:pumping}

As we increase the pumping at a fixed detuning $\Delta$, the effective detuning $\Delta'$ tend to diminish monotonically. However, it is interrupted by the aforementioned resetting of the atomic state, which is subject to the periodic Rabi flopping.  Like other characteristics of the microlaser such as $\langle n \rangle$ or Mandel Q, the frequency pulling also undergoes a periodic change as a function of pumping.

The pumping parameter of the micromaser is given by $\theta=\sqrt{N_{ex}}g\tau$ where $N_{ex}=N/2\gamma_c\tau$. As discussed in Sec.\ \ref{subsec:meaning}, we expect the overall pumping dependence to be $\theta^2$ and its periodic recurrence at the multiple thresholds.
The pumping dependence in the standard micromaser theory \cite{Filipowicz1986, Meystre1988} is usually presented as $g\tau$ is scanned at a fixed atomic flux whereas scanning $N_{ex}$ at fixed $g\tau$ is more relevant to the experimental practice of the microlaser \cite{An1994, FangYen2006}.

\subsubsection{Increasing the atom-cavity coupling}

\begin{figure} \centering
\includegraphics[width=3.4in]{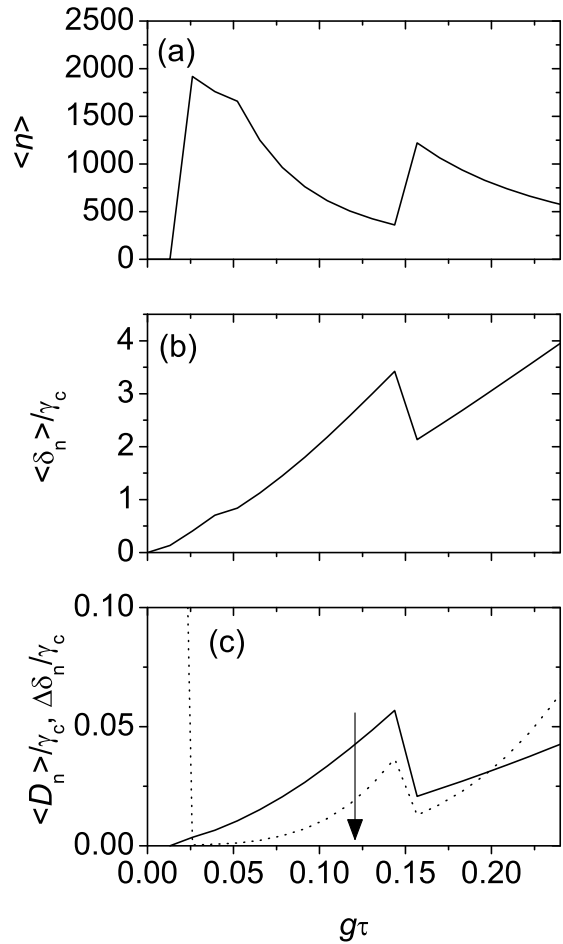}
\caption{The pumping dependence in terms of the atom-cavity coupling constant $g$ at the fixed detuning $\Delta/\gamma_c=-15.0$ and $N=250$. (a) The mean photon number, (b) the mean frequency pulling and (c) the spectral broadening $\langle D_n\rangle/\gamma_c$ (dotted) and $\Delta\delta_n/\gamma_c$ (solid). The arrow indicates the magnitude of the maximum achievable $g$ in the current experimental setting of the microlaser \cite{Choi2006}.} \label{fig:pumping1}
\end{figure}

We tune the coupling constant $g$ at a fixed atom number $N$ in Fig.\ \ref{fig:pumping1}. The multiple thresholds are found for all depicted variables. The mean photon number maintains its typical feature \cite{Filipowicz1986} regardless of the correction by the presence of $\delta_n$. The increase of the frequency pulling in the first branch solution is mildly nonlinear in Fig.\ \ref{fig:pumping1}(b) due to the aforementioned $g^2$ effect. The curvature slightly changes at the first threshold of $g\tau\simeq0.02$. The amount of the pulling induced by a small number of gain atoms is enhanced by the large coupling constant.
The spectral broadening generally becomes larger for stronger pumping except in the multiple-threshold regions, where it drops to a smaller value.
Up to the first branch solution the dispersive broadening dominates over the non-dispersive one.

\subsubsection{Increasing the atomic flux}

\begin{figure} \centering
\includegraphics[width=3.4in]{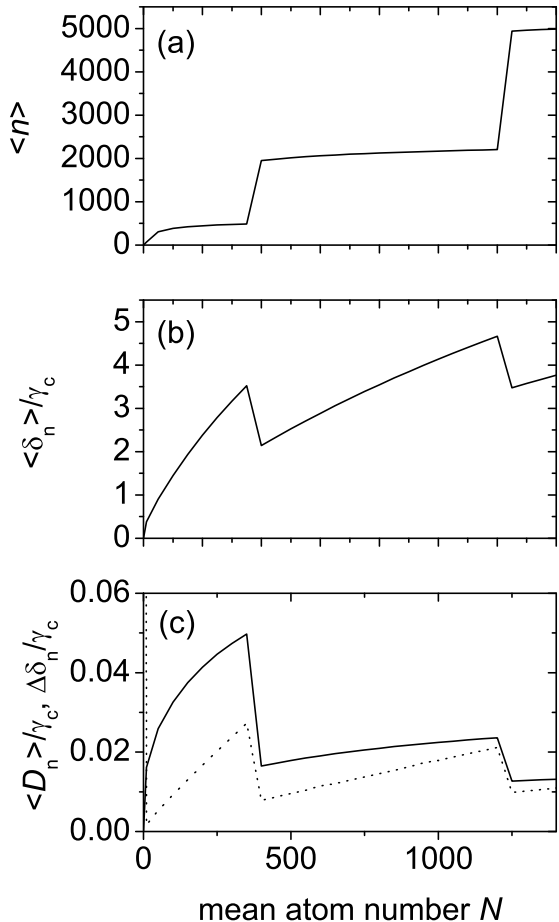}
\caption{The dependence on the pumping via the atomic flux (mean atom number $N$) at the fixed detuning $\Delta/\gamma_c=-15.0$ and $g\tau=0.124$. (a) The mean photon number, (b) the mean frequency pulling and (c) the spectral broadening $\langle D_n\rangle/\gamma_c$ (dotted) and $\Delta\delta_n/\gamma_c$ (solid).} \label{fig:pumping2}
\end{figure}

The pumping by the atomic flux also brings about the multiple thresholds as shown in Fig.\ \ref{fig:pumping2}. The mean photon number is stepwise as verified in the previous experiments \cite{FangYen2006,Hong2009,Seo2010}. The frequency pulling also increases here within each branch solution. It keeps increasing overall except for the periodic drops. The spectral broadening is dominated by the frequency pulling 
for the parameters shown in Fig.\ \ref{fig:pumping2}.

\section{Semiclassical theory}\label{sec:semiclassic}

In this section, we construct a semiclassical theory for intuitive understanding of the frequency pulling phenomena. The source term of the Maxwell equation clarifies the origin of the frequency shift in terms of the Bloch vector. While the semiclassical theory of the conventional laser is comprised of a pair of equations for both intensity and oscillation frequency, the semiclassical equations concerning the microlaser have so far been limited to the intensity alone \cite{An2003, Casagrande1993}. Here we provide the other equation for the frequency shift as well as consequential modification of the intensity equation. The correspondence to the quantum theory is also made in the mean frequency pulling.

\subsection{Maxwell-Schr\"{o}dinger equation}\label{subsec:msequation}

A set of coupled equations describing the microlaser is comprised of the Maxwell equation for the cavity field driven by the induced atomic polarization and the Schr\"{o}dinger equation for the unitary evolution of the atomic state under the electric dipole interaction. In the slowly varying envelope approximation the Maxwell equation reduces to
\begin{equation} {\dot{\cal
E}}(t)+\gamma_c{\cal
E}(t)=i\frac{2\pi\omega_c}{V}\int_{0}^{\tau}{\cal P}(t')\frac{dt'}{\tau},
\end{equation} and the Schr\"{o}dinger equation to
\begin{equation} {\dot
{\cal P}}(t)-i(\omega_c-\omega_0){\cal
P}(t)=-i\frac{\mu^2}{\hbar}{\cal E}(t)r(t),
\end{equation}and
\begin{equation} {\dot
r(t)}=\frac{1}{\hbar}\textrm{ Im}[{\cal
E}^*(t){\cal P}(t)]\end{equation}
where ${\cal E}$ and ${\cal P}$ are the slowly varying amplitudes of the cavity electric field and the induced atomic polarization, respectively, $V$ is the mode volume of the cavity field, $\mu$ is the induced atomic dipole moment and $r$ is the atomic population inversion \cite{Casagrande1993, An2003}. Whereas the previous studies left the variables ${\cal E}$ and ${\cal P}$ complex \cite{An2003, Casagrande1993}, we find it convenient to decompose them into an amplitude-phase form to highlight the role of the in-phase and quadrature components of the induced dipole. With normalization in mind we rewrite
\begin{equation}
{\cal
E}(t)=\sqrt{\frac{V}{8\pi\hbar\omega_c}}a(t)e^{-i\phi(t)},
\end{equation} \begin{equation} {\cal
P}(t)=\mu[S_1(t)-iS_2(t)]e^{-i\phi(t)} \end{equation} where
$a(t), \phi(t), S_1(t)$ and $S_2(t)$
are all real. We rename $r(t)=S_3(t)$ in the context of the Bloch vector. The corresponding Maxwell equation then reads
\begin{equation} {\dot a}(t)+\gamma_c
a(t)=Ng\int_{0}^{\tau}\frac{S_2(t')}{2\tau}dt',\label{aEq} \end{equation}
\begin{equation} {\dot
\phi}(t)=-Ng\int_{0}^{\tau}\frac{S_1(t')}{2a(t')\tau }dt',\label{dEq}
\end{equation}
and the Schr\"{o}dinger equation turns into
\begin{equation} {\dot S_1}(t)-\left(\Delta+{\dot \phi}\right)S_2(t)=0,\label{cEq} \end{equation}
\begin{equation} {\dot S_2}(t)+\left(\Delta+{\dot \phi}\right)S_1(t)=2gaS_3(t),\label{sEq} \end{equation}
\begin{equation} {\dot S_3}(t)=-2gaS_2(t),
\label{rEq}
\end{equation}
where the
atom-cavity coupling constant is given by
$g=\frac{\mu}{\hbar}\sqrt{\frac{2\pi\hbar\omega_c}{V}}$.
Here the time scale involved is different for the field variable and the atomic variables; the atomic state is transient during a single transit through the cavity mode while the field evolves in a longer time scale over many atomic transit events.
We observe that the amplitude of the cavity field is driven by the quadrature component $S_2$ of the atomic dipole while the frequency shift is produced from the in-phase atomic dipole $S_1$. The equations for the atomic variables are exactly the optical Bloch equation except for the shifted field frequency $\Delta'=\Delta+{\dot \phi}$.

With the values of $a$ and $\delta\equiv{\dot \phi}$ held constant in the steady state, we can readily find the solution of the equations of the atomic
variables, Eqs.\ (\ref{cEq})-(\ref{rEq}). The result of this solution will be used to eliminate the atomic variables in Eqs.\ (\ref{aEq}) and (\ref{dEq}) below, arriving at the coupled equation of $a$ and $\delta$ for the determination of their steady state values.

With the initial conditions which represent the perfect population inversion of the incident atoms, $S_3(0)=1$ and $S_1(0)=S_2(0)=0$, the set of solutions is found as
\begin{eqnarray}
S_1(t)&=&\frac{2ga(\Delta+\delta)}{4g^2a^2+(\Delta+\delta)^2} \nonumber\\
&&\times\left[1-\cos\left(\sqrt{4g^2a^2+(\Delta+\delta)^2}t\right)\right],
\end{eqnarray}

\begin{equation} S_2(t)=\frac{2ga}{\sqrt{4g^2a^2+(\Delta+\delta)^2}}\sin\left(\sqrt{4g^2a^2+(\Delta+\delta)^2}t\right),\label{sSol} \end{equation}
 and

\begin{eqnarray}
S_3(t)&=&\frac{1}{4g^2a^2+(\Delta+\delta)^2}\nonumber\\
&&\times\bigl[4g^2a^2\cos\left(\sqrt{4g^2a^2+(\Delta+\delta)^2}t\right) +(\Delta+\delta)^2\bigr].\nonumber\\&&
\end{eqnarray}

\subsection{Equations for intensity and frequency}\label{subsec:ndelta}

The integration of Eq.\ (\ref{aEq}) in the absence of $\delta$ provides the conventionally known gain-loss equation \cite{FangYen2006, Hong2009}. In the generalized approach given here, the dispersive correction to the microlaser gain is made via the effective detuning $\Delta'$. More importantly, we will derive the other equation regarding $\delta$ to complete the dual equations needed for fully describing the intensity and frequency of the microlaser.

The coupled equations can be written in a form for convenient interpretation. First, the integration of Eq.\ (\ref{aEq}) using
Eq.\ (\ref{sSol}) gives
\begin{eqnarray}
{\dot a}(t)+\gamma_c a(t)
&=&\frac{N}{\tau}\left[\frac{2g^2a}{4g^2a^2+(\Delta+\delta)^2}\right]\nonumber\\&&\times\sin^2\left(\sqrt{g^2a^2+(\Delta+\delta)^2/4}\tau\right).\nonumber\\ \end{eqnarray}
Multiplying $2a$ on both sides leads to the
familiar rate equation \begin{eqnarray} {\dot
n}+2\gamma_c n&=&\frac{N}{\tau}\left[\frac{4g^2n}{4g^2n+(\Delta+\delta)^2}\right]\nonumber\\&&\times\sin^2\left(\sqrt{g^2n+(\Delta+\delta)^2/4}\tau\right)\nonumber\\\label{nEq}
\end{eqnarray} where $n=a^2$ (equivalent to $\langle n\rangle$ in the quantum case). Note the modification of the gain-loss description in terms of the effective detuning $\Delta'=\Delta+\delta$. In the steady state (${\dot n}=0$), $n$ is determined by
\begin{eqnarray} n&=&\frac{N}{2\gamma_c \tau}\left[\frac{4g^2n}{4g^2n+(\Delta+\delta)^2}\right]\nonumber\\&&\times\sin^2\left(\sqrt{g^2n+(\Delta+\delta)^2/4}\tau\right)\label{EQ1}.
\end{eqnarray}

\begin{figure} \centering
\includegraphics[angle=0,width=3.2in]{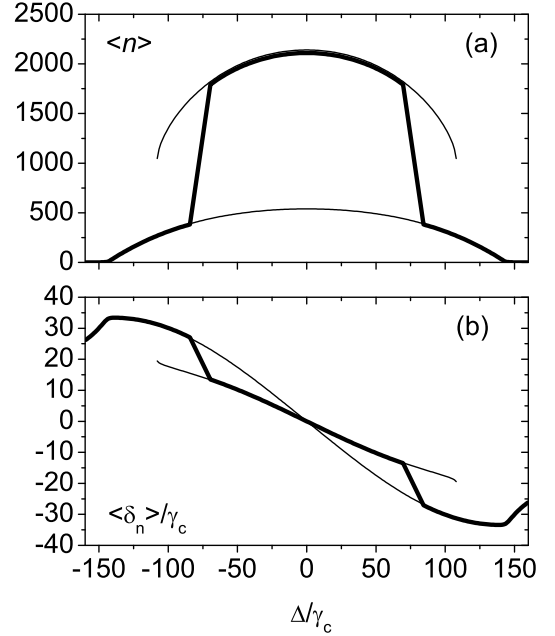}
\caption{The comparison of the quantum (thick curve) and the semiclassical (thin curve) calculations for (a) the mean photon number and (b) the mean frequency pulling. The same parameters are used as in Fig.\ \ref{fig:detuning4}.} \label{fig:correspondence1}
\end{figure}

\begin{figure} \centering
\includegraphics[angle=0,width=3.2in]{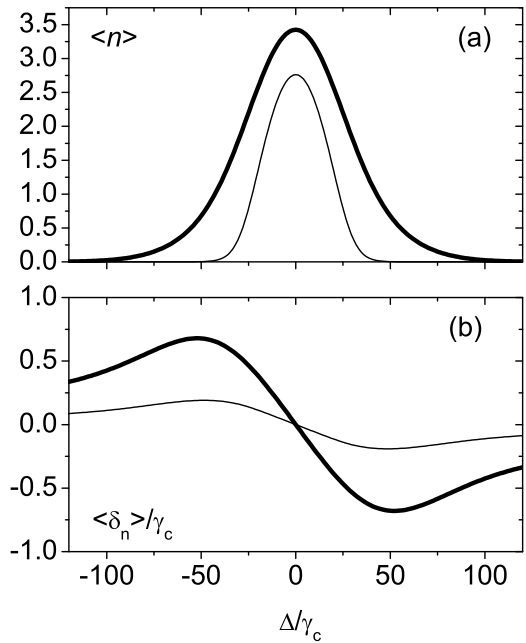}
\caption{The comparison of the quantum (thick curve) and the semiclassical (thin curve) calculations with $g\tau=0.496$ and $N=0.5$ for (a) the mean photon number and (b) the mean frequency pulling.} \label{fig:correspondence2}
\end{figure}

Let us turn to the integration of $\delta$-equation, which is
\begin{eqnarray}
\delta&=&-Ng\int_{0}^{\tau}\frac{S_1(t')}{2a(t')\tau}dt'\nonumber\\
&=&-\frac{Ng^2(\Delta+\delta)}{4g^2n+(\Delta+\delta)^2}\nonumber\\
&&\times\left[1-\frac{\sin\left(\sqrt{4g^2n+(\Delta+\delta)^2}
\tau\right)}{\sqrt{4g^2n+(\Delta+\delta)^2}\tau}\right].
\label{EQ2}
\end{eqnarray}
Note the exact correspondence of Eq.\ (\ref{EQ2}) to the quantum result, Eq.\ (\ref{eq:pulling1}). The two equations, Eqs.\ (\ref{EQ1}) and (\ref{EQ2}), complete the semiclassical description to correctly give the oscillation frequency as well as the intensity. The correction to the previous approach can be easily appreciated by the graphical method presented in Appendix \ref{subsec:graphic}.

Since those coupled equations cannot be separated for individual variables, we presented the numerical calculation of $n$ and $\delta$ in Fig.\ \ref{fig:correspondence1}. They are obtained by iterating the integration until a converging set of $(n, \delta)$ is found. When compared to the quantum calculation of $\langle n\rangle$ and $\langle \delta_n \rangle$, the semiclassical calculation is in good agreement in the example of Fig.\ \ref{fig:correspondence1} and many others.

Of course, the semiclassical theory is limited in the following aspects. First and obviously, the intrinsic statistical properties such as $p_n$ or $\Delta \delta_n$ are not accessible by the semiclassical theory. Second, the semiclassical solution provides only the possible branch solutions if the multiple thresholds are involved. How those multiply segmented branches are connected cannot be traced since it is the quantum phenomena often modeled by the tunneling of the photon number distribution over an effective potential barrier \cite{Filipowicz1986}. Lastly, the possible branch solution itself may not be reliable if one deals with small photon numbers and strong couplings as we encounter in Fig.\ \ref{fig:correspondence2}. In such a regime the distinction between the quantum Rabi frequency $g\sqrt{n+1}$ and the semiclassical approximation $g\sqrt{n}$ becomes significant, and makes a big difference.

\section{Conclusion}

We have formulated a quantum theory of the frequency pulling in the cavity-QED microlaser by considering the effect of the dispersive atom-photon interaction in the off-resonance spectra.
The amount of the pulling induced by a small number of gain atoms turns out to be significant due to the strong atom-cavity coupling, which also gives rise to a distinctive periodic nature in the pumping dependence as contrasted to the conventional laser.
The cavity-tuning curve exhibits a dispersion curve with notable features such as the multiple-step dispersion.
The idea of quantum frequency pulling is introduced and it is shown that the photon-number-dependent frequency pulling induces a spectral broadening, which can be a major source of decoherence in this laser/maser.
The spectral lineshape deviates from a Lorentzian curve depending on the strength of the dispersive interaction involved.
The present work can be used to obtain the theoretical spectrum of the microlaser/micromaser over the entire tuning range.
For the microlaser, a direct measurement of the spectrum is possible owing to the capability of photodetection in the visible region, and such a measurement is expected to arrive soon in view of the recent development of the high sensitivity heterodyne spectroscopy \cite{Hoffges1997, Hong2006, Kim2011}.
The concept introduced here may also be applicable to the other cavity-QED lasers \cite{McKeever2003, Astafiev2007, Dubin2010}.

This work was supported by Korea Research Foundation (Grant Nos. WCU-R32-10045 and 20110015720).

\begin{appendix}

\section{Derivation of $D_n$ and $\delta_n$}\label{subsec:formulae}

We use
\begin{eqnarray}
\Omega_{n+1}&=&\sqrt{4g^2(n+2)+\Delta'^2}=\Omega_n\left(1+\frac{4g^2}{\Omega_n^2}\right)^{1/2}
\nonumber\\&\simeq&\Omega_n+\frac{2g^2}{\Omega_n}
\end{eqnarray}
and
\begin{equation}
\frac{1}{\Omega_{n+1}}\simeq\frac{1}{\Omega_n}-\frac{2g^2}{\Omega^3_n}
\end{equation}
by assuming $n$ much larger than unity. Then the real part of Eq. (\ref{eq:mu_n}) can be written as
\begin{eqnarray}
D_n&\simeq&-r_a\biggl[1+\cos\left(\frac{\Omega_n\tau}{2}\right)\biggl\{\cos\left(\frac{\Omega_n\tau}{2}\right)\cos\left(\frac{g^2\tau}{\Omega_n}\right)\nonumber\\&&-\sin\left(\frac{\Omega_n\tau}{2}\right)\sin\left(\frac{g^2\tau}{\Omega_n}\right)\biggr\}+\left(1-\frac{4g^4}{\Omega_n^4}\right)\nonumber\\&&\times\sin\left(\frac{\Omega_n\tau}{2}\right)\biggl\{\cos\left(\frac{\Omega_n\tau}{2}\right)\sin\left(\frac{g^2\tau}{\Omega_n}\right)\nonumber\\&&+\sin\left(\frac{\Omega_n\tau}{2}\right)\cos\left(\frac{g^2\tau}{\Omega_n}\right)\biggr\}\biggr]-2\gamma_c\sqrt{n(n+1)}\nonumber\\&&+\gamma_c(2n+1)\nonumber\\&\simeq&r_a\biggl[1-\cos\left(\frac{g^2\tau}{\Omega_n} \right)\biggr]+\frac{\gamma_c}{4n}\nonumber\\&=&2r_a\sin^2\left(\frac{g^2\tau}{2\Omega_n} \right)+\frac{\gamma_c}{4n}
\end{eqnarray}
while the imaginary part $\delta_n$ reads
\begin{eqnarray}
\delta_n&=&-r_a\biggl[\frac{\Delta'}{\Omega_{n+1}}\cos\left(\frac{\Omega_n\tau}{2}\right)\sin\left(\frac{\Omega_{n+1}\tau}{2}\right)\nonumber\\&&-\frac{\Delta'}{\Omega_{n}}\cos\left(\frac{\Omega_{n+1}\tau}{2}\right)\sin\left(\frac{\Omega_{n}\tau}{2}\right)\biggr]\nonumber\\
&\simeq&-r_a\biggl[\left(\frac{\Delta'}{\Omega_{n}}-\frac{2g^2\Delta'}{\Omega_n^3}\right)\cos\left(\frac{\Omega_n\tau}{2}\right)\nonumber\\&&\times\biggl\{\frac{g^2\tau}{\Omega_n}\cos\left(\frac{\Omega_{n}\tau}{2}\right)+\sin\left(\frac{\Omega_{n}\tau}{2}\right)\biggr\}\nonumber\\
&&-\frac{\Delta'}{\Omega_{n}}\biggl\{\cos\left(\frac{\Omega_{n}\tau}{2}\right)-\frac{g^2\tau}{\Omega_n}\sin\left(\frac{\Omega_{n}\tau}{2}\right)\biggr\}\nonumber\\&&\times\sin\left(\frac{\Omega_{n}\tau}{2}\right)\biggr]\nonumber\\
&\simeq&-r_a\frac{g^2\Delta'\tau}{\Omega_n^2}\biggl[1-\frac{\sin(\Omega_n\tau)}{\Omega_n\tau} \biggr]
\end{eqnarray}
taking the terms up to the order of $(g/\Omega_n)^2$.

\section{Graphical determination of the steady-state intensity and frequency}\label{subsec:graphic}

\begin{figure} \centering
\includegraphics[angle=0,width=2.63in]{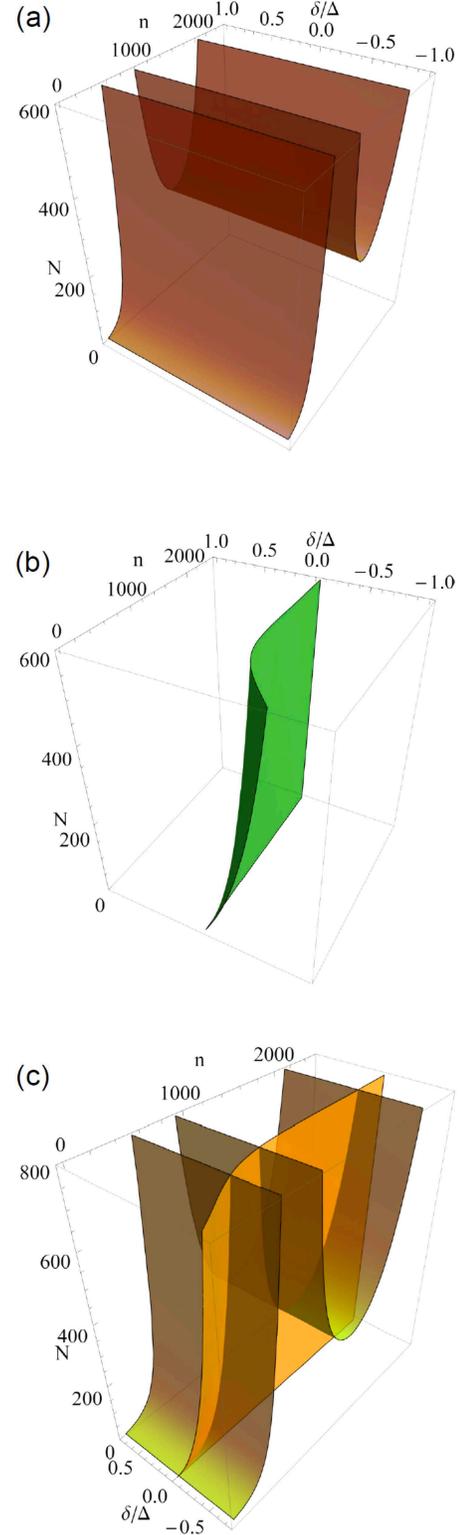}
\caption{(color online) (a) The surface of $N (n, \delta)$ represented by Eq.\ (\ref{eq:N1}). (b) The surface of $N (n, \delta)$ given by Eq.\ (\ref{eq:N2}).  The positive detuning of $\Delta\tau=0.653$ as well as $g\tau=0.124$ and $\gamma_c\tau=0.049$ are used. (c) The solution curve of ($n, \delta$) as a function of $N$ is given by the intersection of the two surfaces in (a) and (b).}
\label{fig:surface}
\end{figure}

To appreciate the effect of the newly found variable $\delta$, let us introduce a graphical method for the determination of the steady-state solution $(n, \delta)$. Equations (\ref{EQ1}) and (\ref{EQ2}) can be rewritten as the mean atom number $N$ represented in $(n, \delta)$-space as
\begin{eqnarray} N&=&\frac{1}{2}\gamma_c \tau\frac{4g^2n+(\Delta+\delta)^2}{g^2}\nonumber\\&&\times\frac{1}{\sin^2\left(\sqrt{4g^2n+(\Delta+\delta)^2}
\tau/2\right)} \label{eq:N1}\end{eqnarray}and
\begin{eqnarray}
N&=&-\frac{\delta}{\delta+\Delta}\times\frac{4g^2n+(\Delta+\delta)^2}{g^2}\times\nonumber\\&&\left(1-\frac{\sin\left(\sqrt{4g^2n+(\Delta+\delta)^2}
\tau\right)}{\sqrt{4g^2n+(\Delta+\delta)^2}\tau}\right)^{-1}, \label{eq:N2}
\end{eqnarray}
respectively. From these two surfaces of $N$ we can construct the evolution of $(n, \delta)$ as a function of the pumping as in Fig.\ \ref{fig:pumping2}.

An example is shown in Fig.\ \ref{fig:surface} for a blue detuning ($\Delta>0$). The surface in Fig.\ \ref{fig:surface}(a) basically represents a multi-branch solution of $n$, here depicted up to the second branch as a function of $N$. The mean photon number or the intensity $n$ does not change much for the range of $\delta$ considered in Fig.\ \ref{fig:surface}(a), in contrast to its marked variation with $N$. Thus the traditionally accepted solution obtained by taking the line of $\delta=0$ is good enough for qualitative understanding of $n$. Determination of the actual nonzero $\delta$ and its corresponding $n$ requires the consideration of Eq.\ (\ref{eq:N2}) as in the surface in Fig.\ \ref{fig:surface}(b). Only at the very bottom of the surface ($N=0$), the solution reduces to $\delta=0$ line. This surface shows the qualitative feature of the pulling. For example, it is bent toward negative $\delta$ as $N$ is grown. The direction of bending is the opposite for red detuning ($\Delta<0$), although not shown.

The solution is then given by the intersection of the two surfaces as shown in Fig.\ \ref{fig:surface}(c).
The dependence of $n$ (or $\delta$) on the pumping via the atomic flux is obtained by
projecting the intersection onto $n$ (or $\delta$) plane as we increase $N$ vertically.
The intensity $n$ does increase as expected, but the bending towards the negative $\delta$ leads to slightly increased $n$ compared to that of $\delta=0$ case. This is understandable since $\delta$ reduced the effective detuning $\Delta'=\Delta+\delta$. Owing to the periodicity in the surface given by Eq.\ (\ref{eq:N1}), the solution is also periodic.

\end{appendix}

\end{document}